\documentclass[%
 reprint,
 showkeys,
 amsmath,amssymb,
 aps,
]{revtex4-1}
\usepackage{placeins}
\usepackage{graphicx}
\usepackage{dcolumn}
\usepackage{bm}

\begin{document}


\title{Static Pressure Extraction from Velocimetry in Vortical Flow Fields}

\author{Nandeesh Hiremath}
\author{Jose C. Sanchez}%
\author{Narayanan Komerath}
\affiliation{%
 Georgia Institute of Technology, School of Aerospace Engineering, Atlanta, Georgia, 30332-015, USA\\
}%

\date{\today}

\begin{abstract}
Rapid fluctuations of static pressure away from flow boundaries in a low-speed vortical flow, are extracted using the measured velocity field. The technique departs from prior work in not requiring far boundary values.  Stereo Particle Image Velocimetry data satisfying the continuity equation are used as input to a Navier-Stokes solver. Streamline curvature, inspired by the Polhamus suction analogy, is used to determine the pressure inside recirculation regions. The technique is validated numerically using the  flow around a circular cylinder before being applied to the vortical flow beneath the sharp edge of a rotor blade in reversed flow. The technique proves effective for sparsely seeded regions including recirculation zones and shear layers in vortices. 
\begin{description}
\item[47.54.De] Experimental aspects
\item[47.32.-y] Vortex dynamics
\end{description}

\end{abstract}

\pacs{Valid PACS appear here}
\maketitle


It is surprisingly difficult to measure static pressure fluctuations in low-speed flows, and more so, away from flow boundaries. The motivating problem here is the pressure field on and near a rotor blade operating in edgewise flow at an advance ratio that is high enough to cause reversed flow on the retreating blade. The sharp edge of the blade encounters forward flow (blunt edge to sharp edge) near the rotor tip, and reversed flow (sharp edge to blunt edge) inboard, crossing zero relative velocity. A strong helical vortex occurs at or near the sharp edge, causing somewhat sharp excursions of lift and pitching moment. The pressure field is as interesting as it is difficult to capture. Intrusive probes are ruled out. Surface-mounted transducers are costly and ineffective on the highly-accelerated blade surface, with low relative air speeds. Pressure Sensitive Paint fails where the relative speed is low. However, low-Mach number flow offers the great simplification that the pressure field can be obtained directly from the velocity field, if the latter is known accurately.

Prior work in this area has focused on two approaches. The first is to differentiate the Navier-Stokes equation and arrive at the Poisson equation, from which pressure is calculated \cite{dabiri2014algorithm, gurka1999computation}. The second uses an iterative marching technique from the boundaries of the computational domain  solution for the elliptical Navier-Stokes differential equation for incompressible inviscid flow. The former technique lacks information about the source/sink terms within the flowfield and is purely dependent on Neumann or Dirichlet boundary conditions. The latter form requires information from all the boundaries with fine spatial resolution to overcome numerical errors.   

In the present approach, a computational tool has been developed, suitable for flowfields far from boundaries. The technique showed promising results for measured velocity fields with limited spatial resolution, limited field of view, and sparsely seeded regions. The design process for the computational tool underwent several iterations using different estimation/correction methods. Initially, the unsteady Bernoulli equation was paired with a discretized velocity field that  was extracted from particle image velocimetry (PIV). The latter served as the computational domain to obtain an initial estimation of the unsteady inviscid static pressure field. A marching computational method was later implemented starting at the known upstream boundary condition. The stagnation pressure of the incoming flow from the upstream boundary was adjusted at every spatial location by integrating the change in stagnation pressure caused by work addition/subtraction evident through the corresponding velocity gradient. Once this was obtained, the static pressure field was extracted simply as the difference between the stagnation pressure and the local dynamic pressure as obtained from the magnitude of the local velocity vector. This work was further extended to include the viscous terms by solving the incompressible 2D Navier Stokes equations. 

Two approaches were used in this method. The first involved viscous terms being added iteratively as correction terms to the assumed Bernoulli pressures in the entire flowfield. The second approach was the aforementioned pressure gradient integration along various paths originating at the known boundaries. Such an approach requires appropriate integration schemes as well as an averaging technique with proper statistical significance. Both of the approaches resulted in similar pressure fields within the uncertainties of the velocity fields.

\begin{figure}[!ht]
	\center
	\includegraphics[width =1\columnwidth]{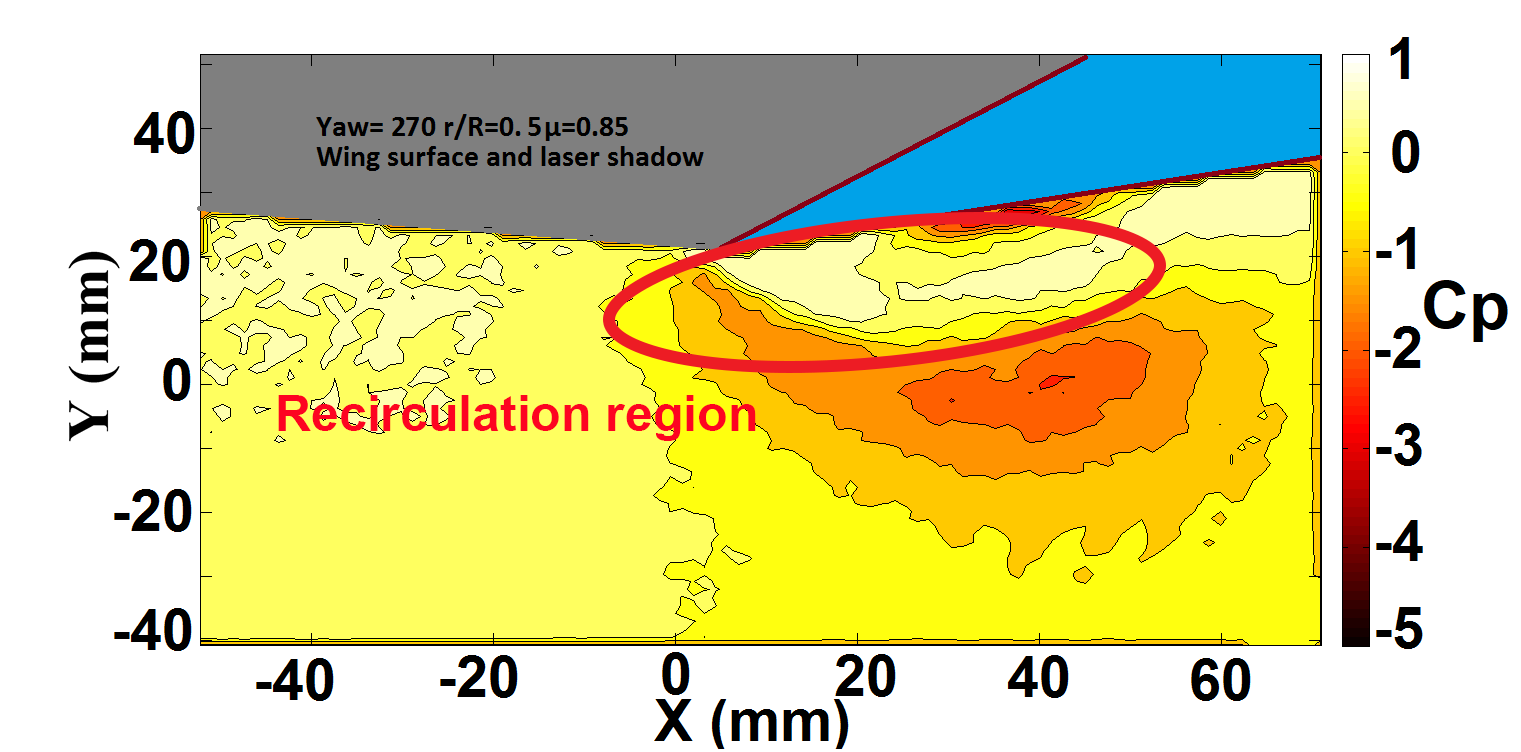}
	\caption{Pressure coefficients showing near stagnation pressures in the recirculation region}
	\label{PressureRecirculaiton}
\end{figure}

The pressure extraction technique was initially implemented for reverse flow regions of rotors near the sharp edge of the retreating blades at high advance ratios ($\mu$). Such flows include a strong helical vortex that develops along the sharp edge. This feature causes first-order effects on the aerodynamics of rotors at high advance ratio. These  regions posed significant challenges for the technique. The field of view of the SPIV fields was comparatively small, consequently the downstream conditions could not be known before hand at the boundary. Consequently, using the  boundary value approach for the elliptic Navier-Stokes equations was not possible. The present technique progresses with only the known upstream and bottom boundary conditions. The second major problem is the absence of velocity vectors in the recirculation region closer to the surface. Although the flow is essentially stagnant in such regions as shown in Figure \ref{PressureRecirculaiton}, the pressure is considerably lower than the freetream stagnation pressure.  A technique based on streamline curvature, inspired by the Polhamus suction analogy, was developed to overcome these challenges. Results were validated with several  independent methods, described further in this letter. 

The first example problem is a circular cylinder in low-Reynolds number cross flow. The Reynolds number is in the regime where Foeppl vortices (closed, attached, symmetric recirculation zones behind the cylinder). This flow has several convenient features. It is steady and laminar. The low Reynolds number means that the effect of the viscous terms in the Navier-Stokes equations are felt throughout the flow. Numerical solutions are obtained quickly, in this case using the COMSOL computational fluid dynamics tool which uses a Reynolds-Averaged Navier-Stokes solver. The closed recirculation zones test the ability of the streamline curvature method. On the other hand, this is a two-dimensional flowfield, enabling swift solution, and with no axial flow in the recirculation zone, which makes it much simpler than the periodic 3D vortex developed in the flow under the rotor blade. 

\begin{figure}[!ht]
	\center
	\includegraphics[width =1\columnwidth]{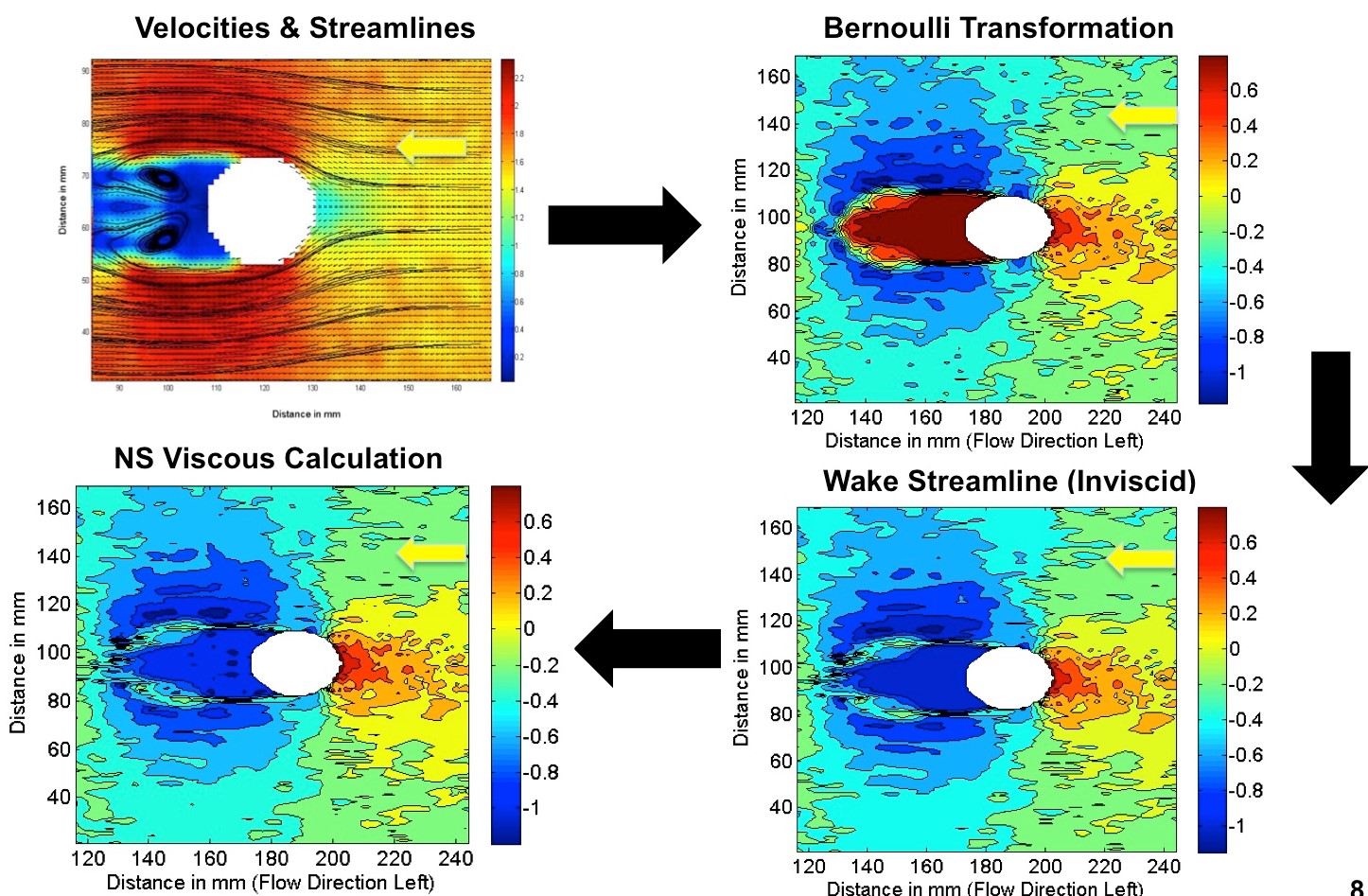}
	\caption{A summary of the test case with the streamline based approach}
	\label{FoepplNumericalDemo}
\end{figure}

\begin{figure}[!ht]
	\center
	\includegraphics[width =1\columnwidth]{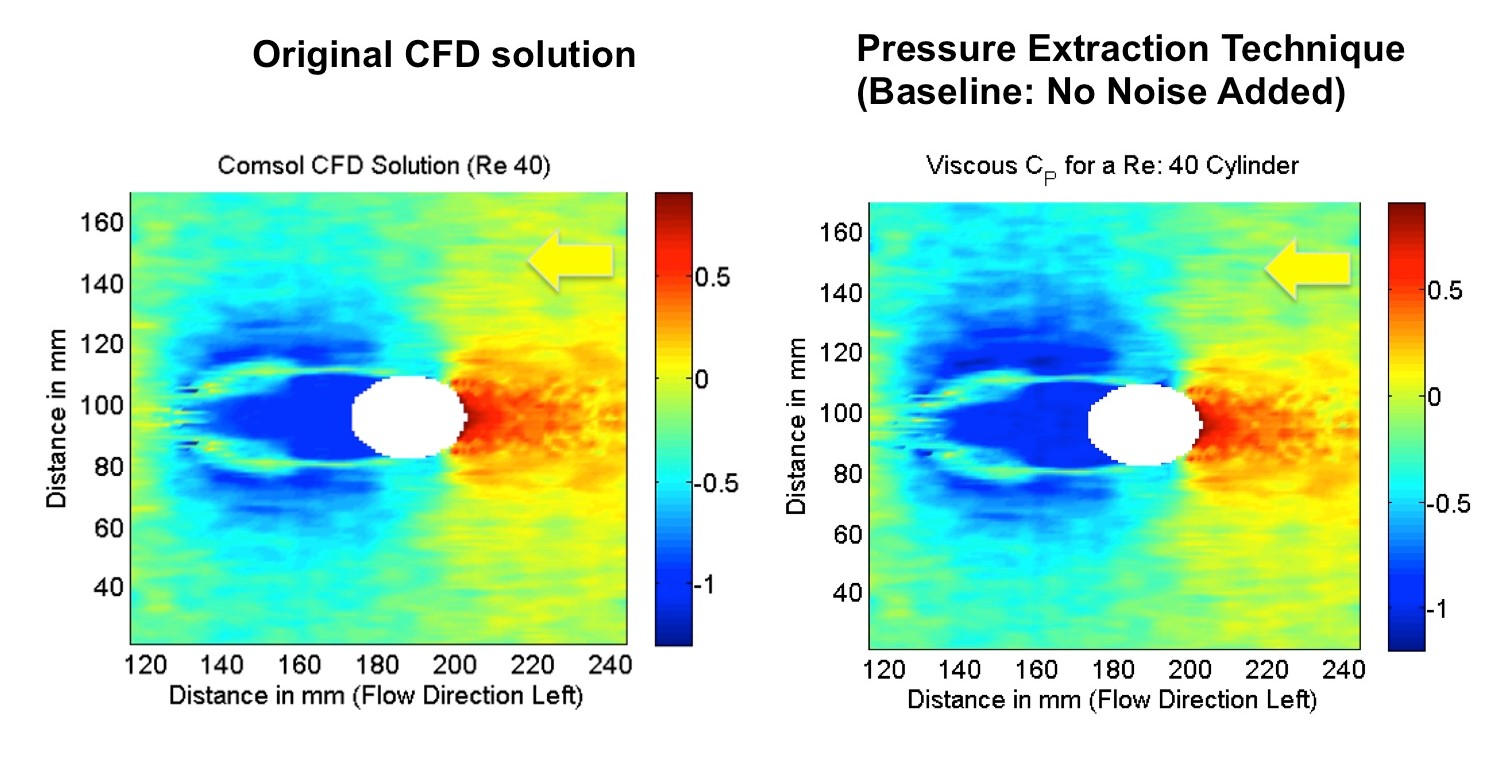}
	\caption{Comparison of the static pressure field extracted from the velocity data, to the actual COMSOL Navier-Stokes result for pressure field of the cylinder}
	\label{FoepplPressureField}
\end{figure}

The velocity field around the 2D cylinder obtained from the COMSOL Navier-Stokes solver was used to obtain the static pressures in the entire flowfield.  Figure \ref{FoepplNumericalDemo} shows  the process involved in the pressure extraction process as well as the streamline correction method and ultimately the viscous corrections. Figure \ref{FoepplPressureField} shows the comparison of the pressure coefficients from the velocity field and the COMSOL Navier Stokes solver. The results show a good agreement in the recirculation region. In this case no random noise has been added to the velocity field to simulate experimental uncertainty. 

Two different integration paths were used to obtain the static pressures, and their effects compared \cite{merkl2015extracting}. The first marched from top to bottom and left to right, while the other marched from bottom to top and right to left. A simple marching scheme from upstream to downstream accumulated large numerical errors; however, the combined integration scheme resulted in smaller discrepancies in the pressure field. To select the most appropriate value, the integration paths were polled by using a median filter by taking the median values of the static pressures from different integration paths leading to a point. Median filters have proven to be more appropriate than simple averaging as shown by Dabiri \cite{dabiri2014algorithm}. 

\begin{figure}[!ht]
	\center
	\includegraphics[width =1\columnwidth]{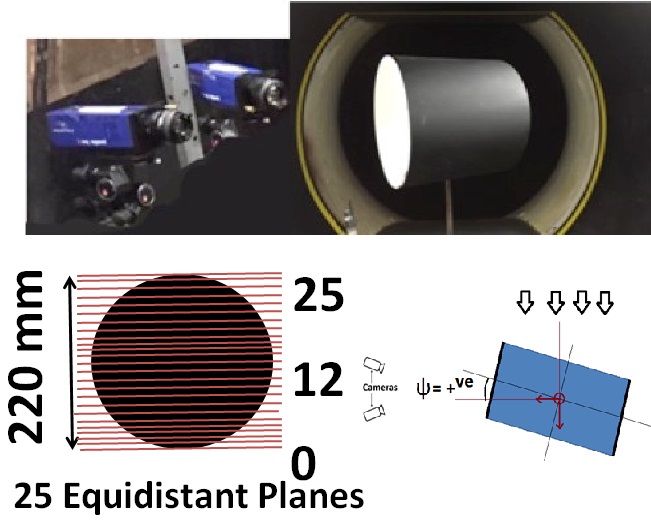}
	\caption{A view of SPIV setup on aspect ratio 1 cylinder, facing upstream}
	\label{SPIVCylSetup}
\end{figure}

The second validation technique was to experimentally measure the surface pressure distribution on the flat surface of a cylinder in yawed cross-flow using pressure transducers. The results were then crosschecked with the near-body pressure distribution obtained from the pressure extraction tool. The wind tunnel experiments were performed at the John J. Harper wind tunnel at our laboratory, focusing on stereo particle image velocimetry (SPIV) and pressure transducer measurements. The cylinder had an aspect ratio of 1, with a 220 mm diameter. It was mounted on a stepper motor to achieve different yaw angles. A maximum blockage of 3-4\% is expected for this setup, alleviated by the vents in the atmospheric-pressure test section. The SPIV cameras were mounted on a traverse with the flexibility to traverse upwards for different sectional planes on the side surface of the cylinder. The experimental setup for Stereo Particle Image Velocimetry (SPIV) is shown in Figure \ref{SPIVCylSetup}. Twenty-five equidistant horizontal planes were selected, parallel to and including the horizontal diameter of the cylinder. The cylinder was placed at yaw settings of $\pm0, \pm4, \pm8, \pm12, \pm20, \pm45,$. Beyond 8 degrees, symmetry was assumed and only the upper 11 planes were used. At each plane, data acquisition had a duration of 30 seconds in order to capture the lowest expected frequencies.

\begin{figure}[!ht]
	\center
	\includegraphics[width =0.9\columnwidth]{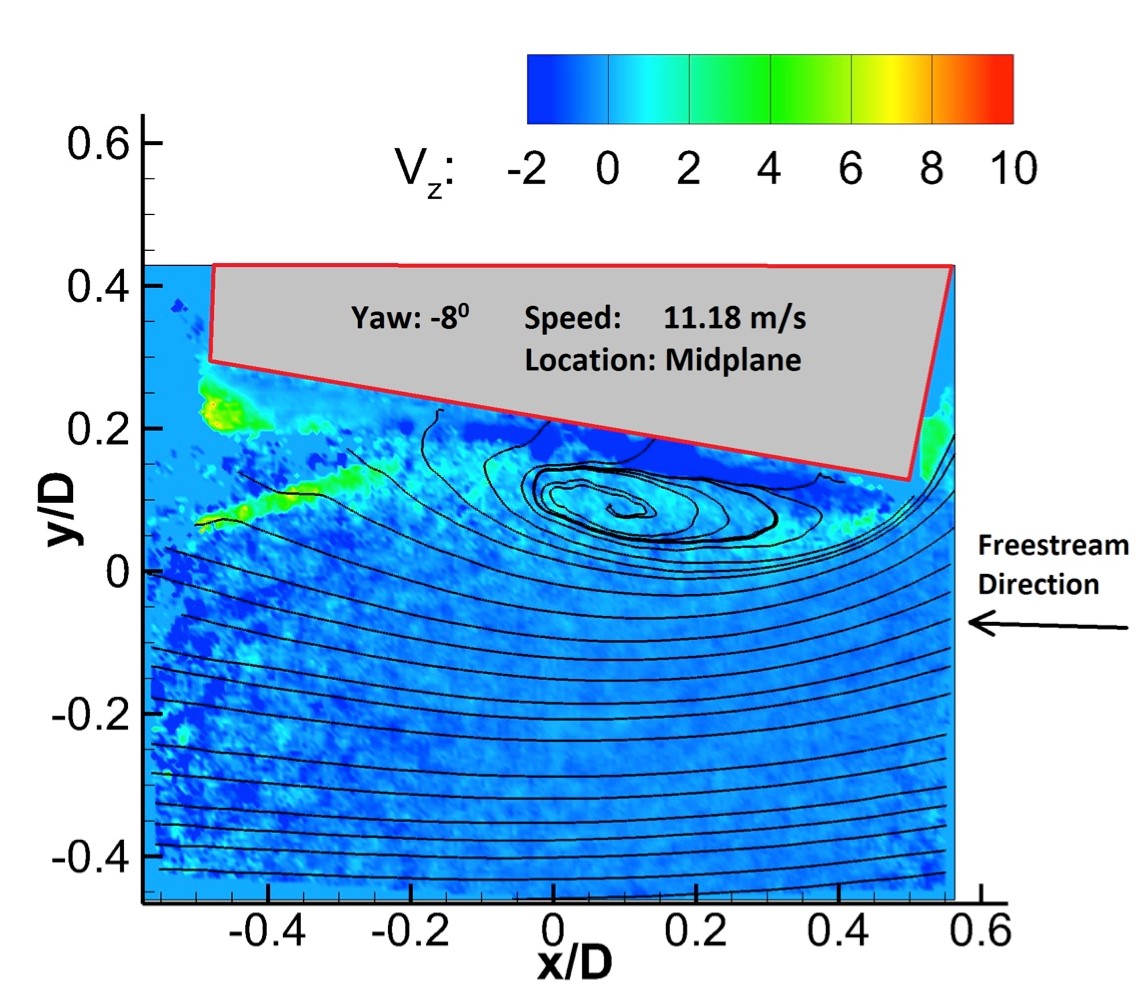}
	\caption{Leeside side streamlines and out-of-plane velocity at the mid-plane of the AR1 cylinder at 8 degrees yaw and 11.8 m/s.}
	\label{SPIVmin8_11}
\end{figure}

Dividing the data acquisition into parallel equidistant planes allows for the accurate retrieval of all three components of velocity \cite{hiremath2017}. At the low and negative yaw cases on the leeward flat face of the cylinder, the velocity field can be easily converted to static pressure because viscous losses are low, demonstrated by the quick convergence of the iterative viscous correction method. For the suction region occurring at high yaw angles, the full Navier-Stokes calculation was performed by iteration to solve for the pressure field. Figure \ref{SPIVmin8_11} shows an example result at -8 degrees yaw (suction side), at the horizontal diameter plane. The freestream is from right to left, at 11.8 m/s. A large vortical structure is apparent, followed by reattachment downstream. The color bar denotes the out-of-plane velocity component. 

\begin{figure}[!ht]
	\center
	\includegraphics[width =1\columnwidth]{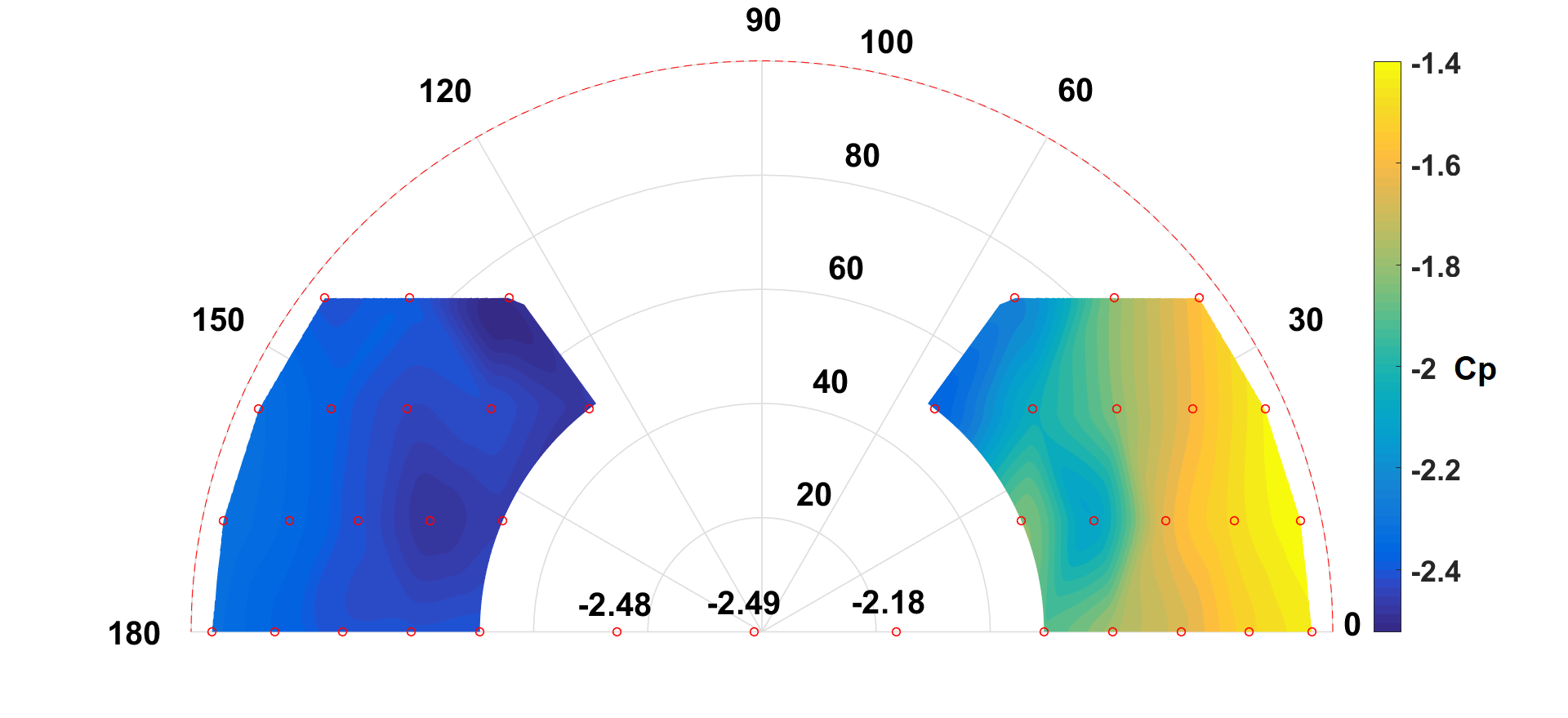}
	\caption{Static pressure measurements from pressure transducer on leeward side at $\Psi = 8^{\circ}$ and $U_{\infty} = 11.17 m/s$}
	\label{Leeward_az8_1117mps}
\end{figure}

The static pressure in the separated region following the reattachment point is also a subject of study for this case. Figure \ref{Leeward_az8_1117mps} shows the pressure transducer port locations on the side surface, which were placed at sufficient resolution to capture the leading edge suction, followed by few points on the mid plane.

\begin{figure}[!ht]
	\center
	\includegraphics[width =1\columnwidth]{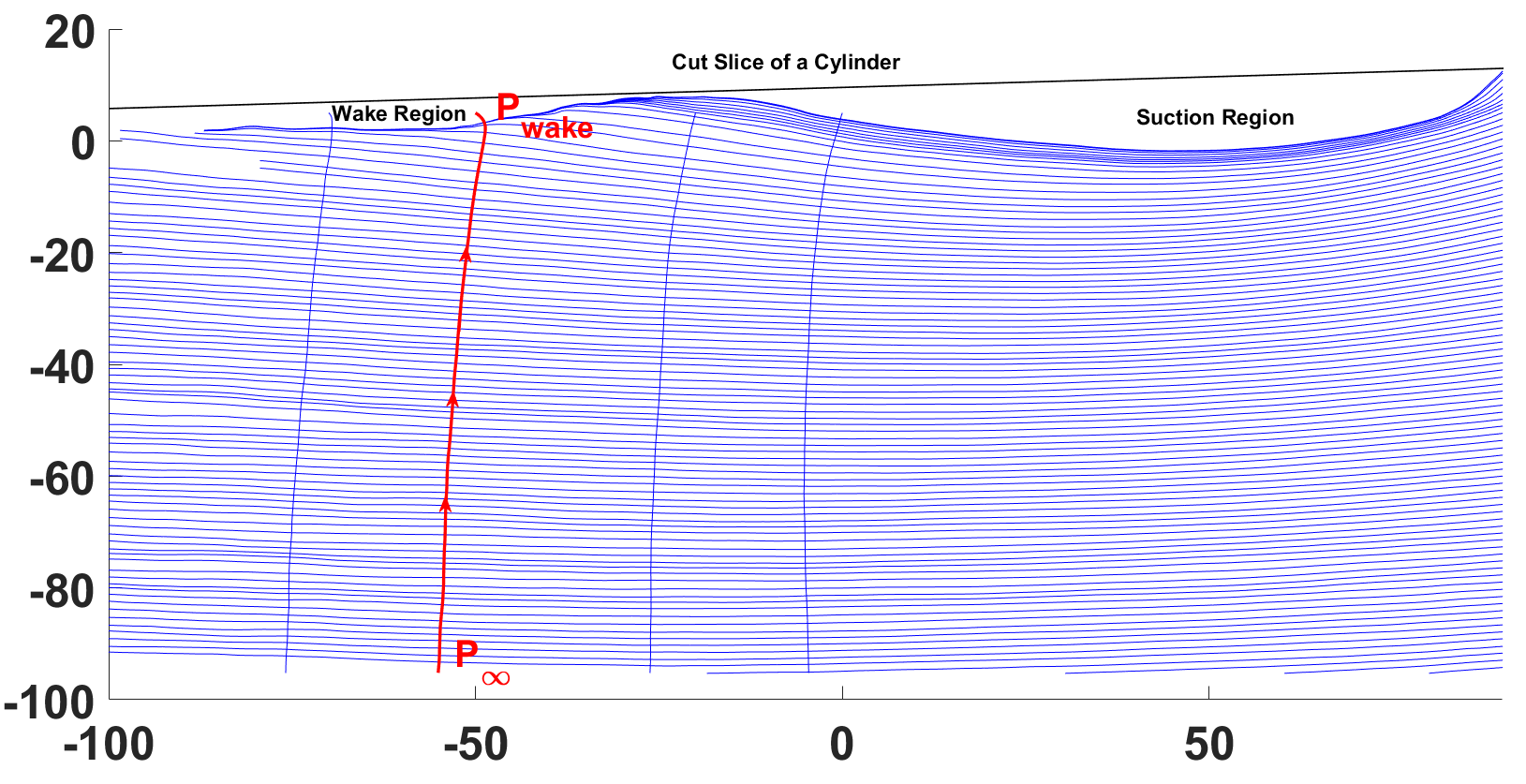}
	\caption{Integration of pressure gradient along potential line}
	\label{StreamlineCurvature}
\end{figure}

As mentioned before, reverse flow data sets had unpopulated velocity vectors in the recirculation region. Consequently a streamline based approach had to be implemented by estimating the pressures from the neighboring streamlines around the recirculation region. This streamline approach follows the Polhamus suction analogy, by determining the curvature of the streamlines as a measure of the pressure gradient. In the case of a delta wing, the high suction pressure at the sharp edge creates a curvature in the streamlines resulting in flow turning and development of a leading edge vortex \cite{37Polhamus1966}. The streamline approach was used to determine the near-field pressure along the bounding shear layer. To obtain the values of the static pressure distribution in the wake region, first the boundary condition at the lower boundary was obtained from a Bernoulli calculation. Subsequently, streamlines along a potential line which terminates within the wake region were selected as a basis for the pressure gradient marching solution based on streamline curvature. The calculation was performed with the equation below, where R is the radius of curvature of the streamline, and U is the local velocity of points along the direction $n$. The pressure gradient was integrated along the potential line for each available streamline until a pressure value was obtained for the specific region in the wake. This methodology proves to be suitable for correcting the erroneous near stagnation pressures in the wake regions. This is verified using  the surface pressures obtained from the pressure transducers on the cylinder. Figure \ref{StreamlineCurvature} represents how an   initial guess for pressures in the wake region is obtained.

\[\frac{dP}{dn} = \frac{\rho U^{2}}{R}\]

\begin{figure}[!ht]
	\center
	\includegraphics[width =1\columnwidth]{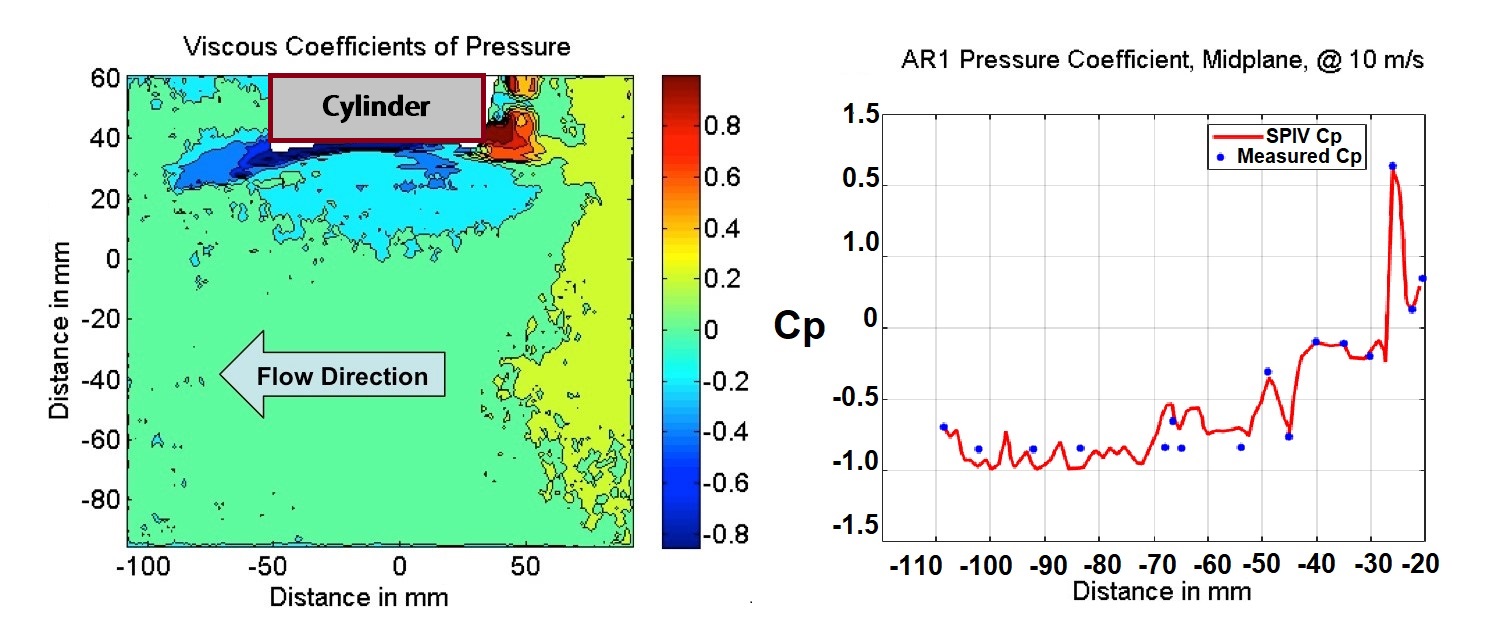}
	\caption{Comparison of pressure coefficients from SPIV field with the pressure transducer data along the mid plane of aspect ratio one cylinder}
	\label{PrQuantVal}
\end{figure}

The data were validated with the pressures extracted from the pressure transducers shown in Figure \ref{Leeward_az8_1117mps} of the top half of the cylinder in the corresponding yaw configuration. The pressure field was interpolated with a spline function from the red discretized pressure transducer locations. The pressure fields along this mid-plane were directly compared to the results obtained from the computational tool as shown in Figure \ref{PrQuantVal}. The two curves show great agreement along the diameter of the cylinder for both the suction region and the wake. The ``jaggedness" in the computational tool's data set in the wake region is likely due to numerical errors from the integration and uncertainty in the velocity measurements. 

\subsection{Flow Under a Retreating Rotor Blade in Reverse Flow} 
The stagnation pressure loss inside the reverse flow vortex due to the viscous losses, was studied to get an order of magnitude of the viscous terms in the Navier-Stokes equation. The starting point was the mechanism for entropy change along a streamline as formulated in the following equations. The entropy ($S*$) along the streamline (s) at temperature (T) and specific volume (v) is dictated by the velocity (U) gradient, the total enthalpy ($h_{\circ}$), and the total pressure ($P_\circ$). With the calorically perfect gas assumption in an incompressible flow field, the change in total enthalpy is zero. This may not be true at the rotor disc itself due the work addition/subtraction. The stagnation pressure loss can then be directly related to the velocity gradient along a streamline.  

\[T\frac{dS*}{ds} = \int \mu (\frac{\partial U}{\partial s})^{2} ds\  =\ \rho v (\int \frac{dh_{\circ}}{ds} - \frac{1}{\rho}\frac{dP_{\circ}}{ds})\]

The entropy relations show that along a streamline due to the viscous effects, stagnation pressure loss scales with the second power of the velocity gradient for an adiabatic flow. The gradient term is observed to be of the order of one with viscosity of air as a multiplicative factor, thus making it negligibly small, of the order of $10^{-5}$.

\begin{figure}[!ht]
	\center
	\includegraphics[width =1\columnwidth]{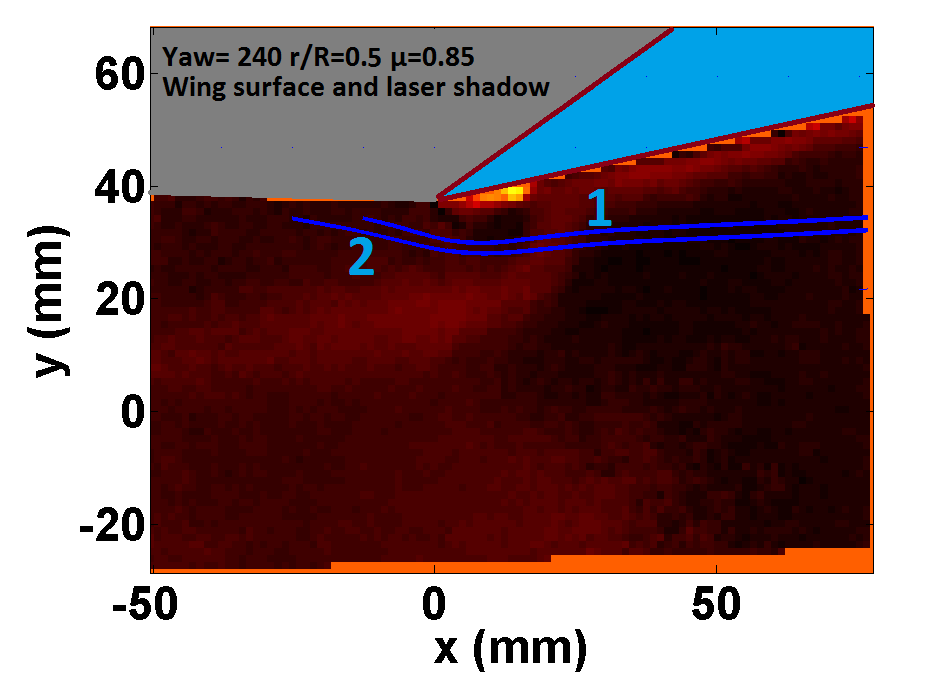}
	\caption{Streamlines in the vicinity of vortex}
	\label{Streamtube}
\end{figure}

\begin{figure}[!ht]
	\center
	\includegraphics[width =1\columnwidth]{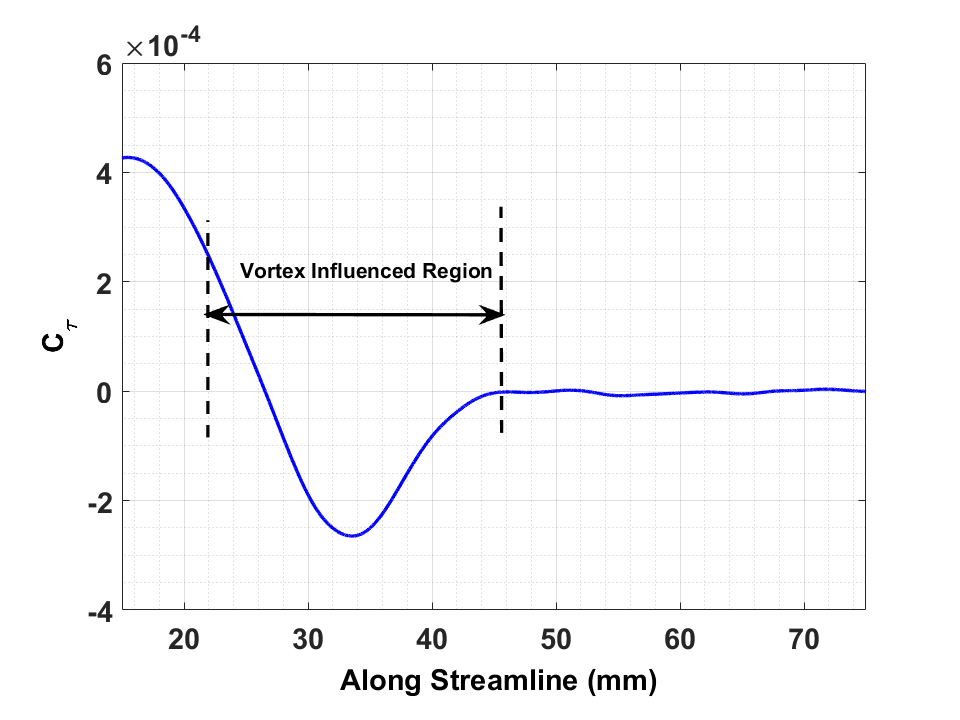}
	\caption{Shear stress coefficient along streamline as a measure of viscous dissipation and stagnation pressure loss along streamline-2}
	\label{ShearStressInsideVortex}
\end{figure}

A Couette flow analogy was used to obtain the actual magnitude of the viscous stresses in the vortex dominated flow field. Figure \ref{Streamtube} shows two streamlines in the vicinity of the vortex. The radius of curvature of the vortex is large compared to the distance between these streamlines, so that a Couette flow problem is approximated well. The analogous wall friction for a Couette flow was obtained at discrete locations on the streamline. Figure \ref{ShearStressInsideVortex} follows the discussion on pressure loss measured by the Couette flow analogy. The shear stress coefficient, just like the skin friction coefficient in the Couette flow case, conveys information about the stagnation pressure loss due to friction as compared to dynamic pressure. Pressure losses along streamline-2 shown in Figure \ref{Streamtube}, measured as an artifact of shear stress, attains negative values in the vortex influenced regions and goes to zero as the streamline leaves the vortex. Pressure loss as compared to freestream dynamic pressure is of the order of $10^{-4}$ through this small region of interest. These low frictional values as compared to convective terms in the Navier-Stokes equation show the signs of very low viscous losses. This finding shows a greater potential in simplifying the complex flow field as approximately an inviscid flow.  \index{Couette flow} \index{Stagnation pressure loss}

\begin{table}[!ht]
	\centering
	\caption{Uncertainty estimates in SPIV and pressure computations}
	\begin{tabular}{ |c||c| }
		\hline
		Parameter  & Error\\
		\hline
		In-plane random error & 0.088 - 0.281 pixels\\
		In-plane velocity error($\varepsilon_u$, $\varepsilon_v$ )& 0.006 m/s - 0.02 m/s \\
		Out-of plane velocity error ($\varepsilon_w$) & 0.028 m/s - 0.281 m/s \\
		Total measurement error & 0.75\% - 3.96\% \\
		Numerical integration grid size & 1.2 mm - 1.6 mm\\
		Error in static pressure ($\Delta P$)& 0.18 - 0.6 Pa\\
		\hline
	\end{tabular}
	
	\label{Tab:UncertaintyEstimates}
\end{table}

\begin{figure}[!t]
	\center
	\includegraphics[width =1\columnwidth]{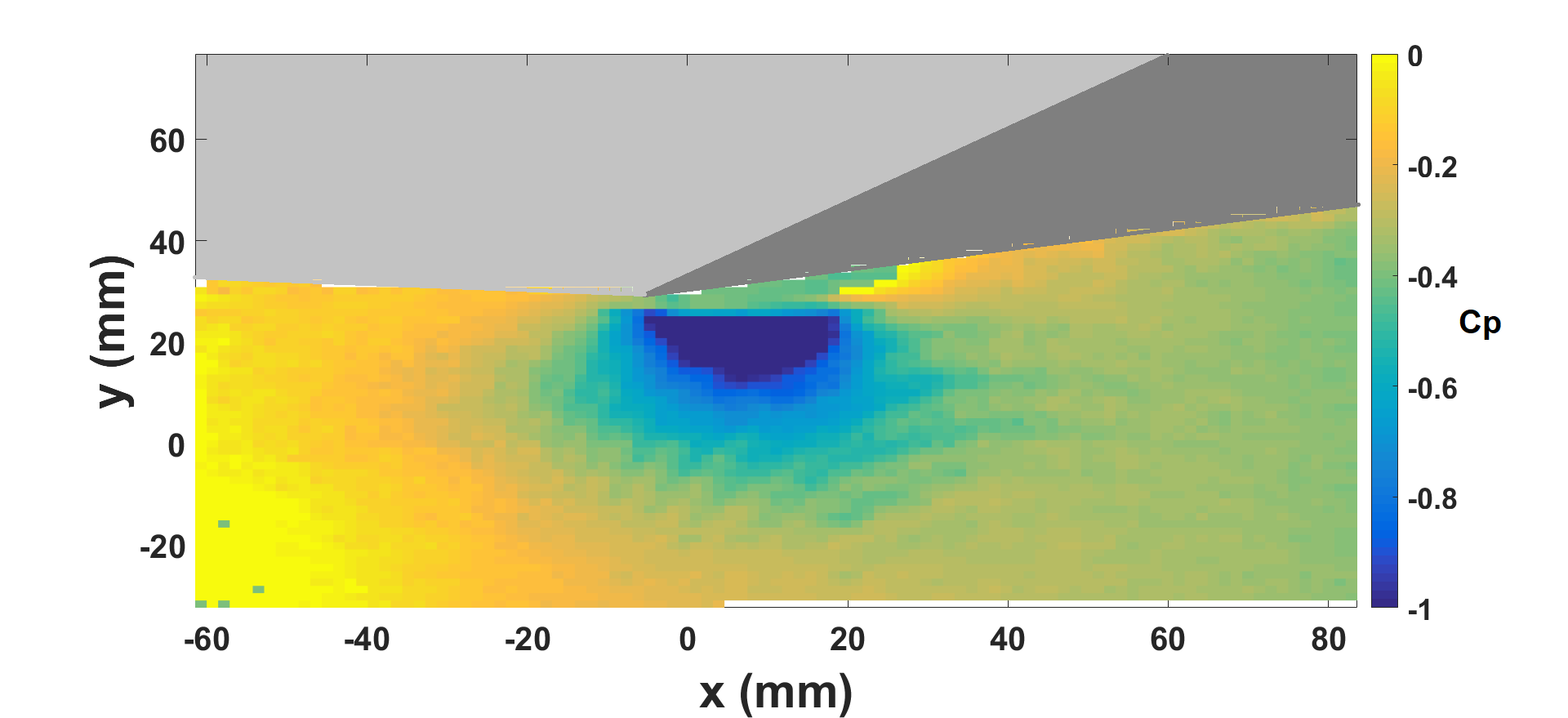}
	\caption{Pressure field at $\Psi = 240^{\circ}, \mu = 0.85, r/R = 0.4 $}
	\label{PrAz240rR40mu85}
\end{figure}

\begin{figure}[!ht]
	\center
	\includegraphics[width =1\columnwidth]{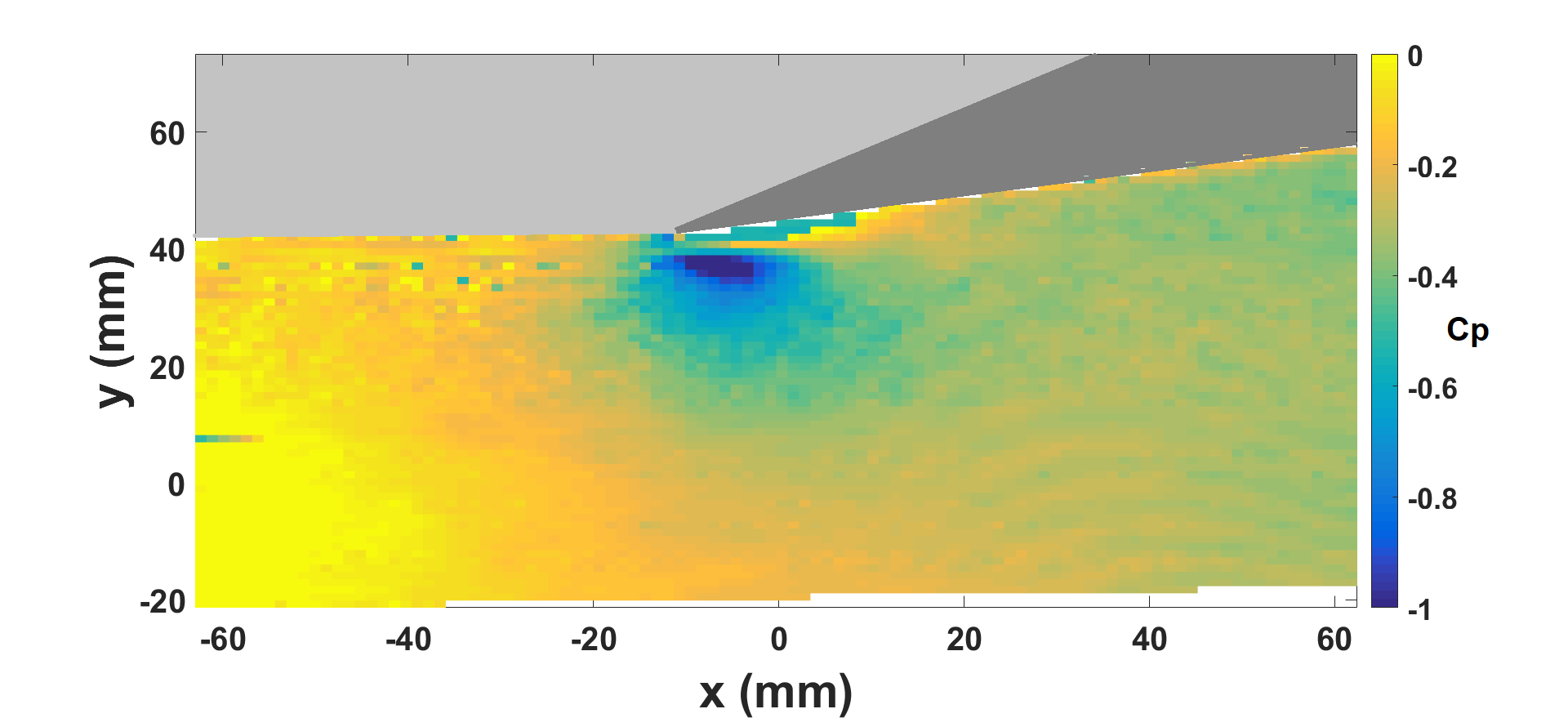}
	\caption{Pressure field at $\Psi = 240^{\circ}, \mu = 0.85, r/R = 0.514 $}
	\label{PrAz240rR514mu85}
\end{figure}

Once validated, the computational tool was employed for SPIV data sets of rotors in the reverse flow region at high advance ratios to explore the nature of the sharp edge vortex. Multiple azimuth angles at different radial locations and advance ratios were examined. The pressure-velocity gradient relationship also helps explore the relative strength of the sharp edge vortex at different radial stations as it grows and eventually bursts and dissipates. Presented here in Figures \ref{PrAz240rR40mu85} and \ref{PrAz240rR514mu85} are the pressure fields at azimuth 240, advance ratio ($\mu$) 0.85, at various radial locations. These individual radial stations can be combined to obtain the surface pressure distribution over the blade, which could provide an indirect load measurement on the rotor.

The work described in this brief Letter is mostly 2-dimensional, with only one SPIV plot shown where the out-of-plane velocity component ($V_z$) is mentioned. Extending to 3D involves adding a Continuity Equation solver and interpolating the grid to the levels required for high-resolution computations. Only the in-plane velocity components are considered in Table \ref{Tab:UncertaintyEstimates} for static pressure uncertainty. The numerical integration grid size is nothing but the pixel size. A grid independence check, or sub pixel interpolation are not considered in this letter.

In conclusion, this Letter describes a technique to extract static pressure in an unsteady flow from velocity data, even where the undisturbed boundary values are not accessible. The method of streamline curvature based on the Polhamus Suction Analogy, permitted a breakthrough in dealing with separated, recirculation zones where the stagnation pressure is unknown. Extension to include turbulent Reynolds stresses is straigthtforward in concept, albeit requiring large storage (the full 3-component velocity data must be stored, and they must be acquired at sufficient temporal rate to span the turbulence spectrum) and additional computations. 


\FloatBarrier
\bibliography{apssamp}

\end{document}